\documentclass[aps,prx,a4paper,notitlepage,reprint]{revtex4-2}

\usepackage{amsmath,amssymb,amsfonts} 
\usepackage{mathrsfs} 

\usepackage{mathtools} 
\usepackage{color}

\usepackage{graphicx,float}
\usepackage[colorlinks,
linkcolor=red,
citecolor=blue,
urlcolor=red]{hyperref}

\newcommand{\ket}[1]{\vert{#1}\rangle}
\newcommand{\bra}[1]{\langle{#1}\vert}
\newcommand{\avg}[1]{\left\langle{#1}\right\rangle}

\newcommand{\re}{\mathrm{Re}\,}
\newcommand{\im}{\mathrm{Im}\,}
\newcommand{\abs}[1]{\left\vert{#1}\right\vert}

\renewcommand{\t}[1]{\mathrm{#1}}

\newcommand{\figref}[1]{Fig.~\ref{#1}}
\renewcommand{\eqref}[1]{Eq.~(\ref{#1})}
\newcommand{\secref}[1]{Sec.~\ref{#1}}

\begin{document}
\title{Prediction-retrodiction measurements for teleportation and conditional state transfer}

\author{Sergey A. Fedorov}

\author{Emil Zeuthen}
\email{zeuthen@nbi.ku.dk}

\affiliation{Niels Bohr Institute, University of Copenhagen, Copenhagen, Denmark}

\begin{abstract}
Regular measurements allow predicting the future and retrodicting the past of quantum systems. Time-non-local measurements can leave the future and the past uncertain, yet establish a relation between them. We show that continuous time-non-local measurements can be used to transfer a quantum state via teleportation or direct transmission. Considering two oscillators probed by traveling fields, we analytically identify strategies for performing the state transfer perfectly across a wide range of linear oscillator-field interactions beyond the pure beamsplitter and two-mode-squeezing types.
\end{abstract}

\maketitle



Classically, specifying the parameters of a closed system at one time allows inferring their values in the past and the future. No information can be added by making new measurements at later times or revealing the outcomes of measurements made before. In quantum mechanics, successive observations of a system completely specified at one time can nevertheless add new information~\cite{aharonov_is_1984}. Such situations were first considered in the context of relativistic quantum theory~\cite{aharonov_time_1964, aharonov_is_1984}, 
where it was argued that the notion of a quantum state has to be extended in order to logically describe systems between measurements, and, to this end, multi-time states were introduced~\cite{aharonov_two-state_2008,aharonov_multiple-time_2009}. Later, these ideas were extended to open systems, and became instrumental in deriving the general statistical theory of past observations~\cite{barnett_bayes_2000,tsang_optimal_2009,gammelmark_past_2013,guevara_quantum_2015}.
Recently, these theories were applied to optical homodyne records for improving the signal-to-noise ratio in sensing~\cite{bao_spin_2020,bao_retrodiction_2020}, and verifying quantum trajectories~\cite{rossi_observing_2019}.

\begin{figure}
    \centering
    \includegraphics[width=\columnwidth]{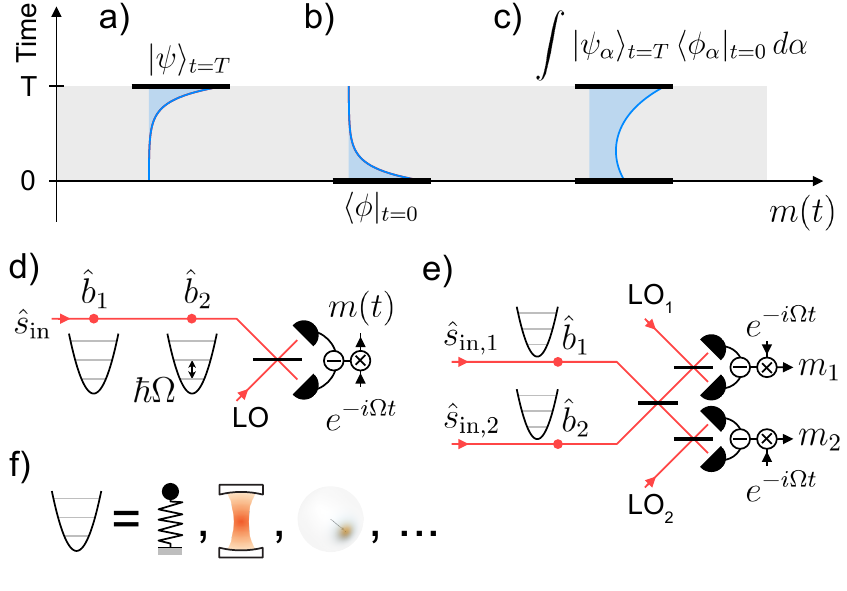}
    \caption{\small{a-c) Preparations based on measurement records $m(t)$ of a) a forward-evolving quantum state by prediction measurement, b) a backward-evolving state by retrodiction measurement, and c) a two-time state by pretrodiction measurement. d-e) The interaction configurations considered in this work, d) sequential and e) parallel. Traveling electromagnetic fields (red lines) interact with localized oscillators 1 and 2 (red dots) via Hamiltonians~(\ref{eq:mainHam}), and result in homodyne measurement currents $m(t)$ and $m_{1,2}(t)$. LO: local oscillator. All beamsplitters are 50:50.
    f) The localized oscillators in the schemes can be realized by mechanical resonators, optical cavities, collective spins, etc.}}
    \label{fig:QM-tasks}
\end{figure}

To date, the analysis of prediction and retrodiction, whether involving multi-time~\cite{aharonov_two-state_2008} or past quantum states~\cite{gammelmark_past_2013}, concentrated on the statistics of outcomes during the measurement interval, $[0,T]$.
Here, we consider their preparative aspects, i.e., effects on monitored systems that measurements leave beyond the measurement interval. 
As illustrated in Fig.~\ref{fig:QM-tasks}a-b, prediction prepares usual forward-evolving quantum states $\ket{\psi}_{t=T}$, for which the future is determined and the past is unknown, whereas retrodiction prepares backward-evolving states $\bra{\phi}_{t=0}$, for which the past is known, but the future is not~\cite{aharonov_two-state_2008}. When states of both types are prepared by a single sequence of measurements, they form a product two-time state $\ket{\psi}_{t=T}\,\bra{\phi}_{t=0}$~\cite{aharonov_multiple-time_2009}. Yet another possibility, shown in Fig.~\ref{fig:QM-tasks}c, is the preparation of superpositions of initial and final states---superposition two-time states $\int\ket{\psi_\alpha}_{t=T}\,\bra{\phi_\alpha}_{t=0}\,d\alpha$~\cite{aharonov_multiple-time_2009}. We will refer to measurements that accomplish this as \emph{pretrodiction}.

The concept of pretrodiction measurements enables new analytical insights into the problem of conditional state transfer between quantum systems. 
We demonstrate this for localized harmonic oscillators continuously probed by traveling electromagnetic fields, deriving results that go beyond and complement those of previous analyses based on temporal modes~\cite{hammerer_teleportation_2005,hofer_quantum_2011,hofer_entanglement-enhanced_2015,navarathna_continuous_2023,krauter_deterministic_2013}, Kalman filtering~\cite{muller-ebhardt_quantum-state_2009,wieczorek_optimal_2015}, and path integrals~\cite{khalili_preparing_2010}.
We consider general linear interactions between the oscillators and fields, thereby encompassing optical cavities, optomechanical devices, macroscopic spin ensembles, and microwave resonances in superconducting circuits.
In the first configuration, shown in \figref{fig:QM-tasks}d, one field interacts sequentially with two oscillators over the time $t\in[0,T]$ and is measured by a homodyne detector. In the second configuration, shown in \figref{fig:QM-tasks}e, two independent fields interact with two oscillators in parallel, are combined on a 50:50 beamsplitter, and then measured by two homodyne detectors. 
The measurements are much slower than the oscillation period in both cases, and the homodyne local oscillators have the same carrier frequency as the driving field, meaning that the signals on the photodetectors are concentrated around the common resonance frequency $\Omega$. 

The state transfer from oscillator 2 to 1 can in both setups be accomplished via teleportation, i.e., in spite of no quantum field traveling from 2 to 1. In the sequential scheme in \figref{fig:QM-tasks}d, it is also possible to transfer a state from 1 to 2 via direct transmission, which can be conditional on the homodyne measurements. We find that all these types of transfer can be realized perfectly for a wide range of field-oscillator interactions, even when the input traveling fields have non-zero thermal population.
The latter fact generalizes a known result for electromagnetic cavities~\cite{jahne_high-fidelity_2007,yin_catch_2013,kiilerich_input-output_2019,vermersch_quantum_2017,magnard_microwave_2020}.

\paragraph*{Teleportation via time-non-local measurements ---}
The teleportation of a quantum state \cite{bennett_teleporting_1993,vaidman_teleportation_1994,braunstein_teleportation_1998} is traditionally seen as a stepwise protocol of first creating entanglement between Alice and Bob (corresponding to the interaction between the traveling field and oscillator 1 in our schemes), then measuring the states of Alice and Charlie in the EPR basis (the interaction of the field with oscillator 2 followed by homodyne detection in our schemes), and finally applying feedback to the system of Bob. Vaidman was the first to notice the relation of teleportation to time-non-local measurements~\cite{vaidman_teleportation_1994}. 
We formulate the task of teleportation of an oscillator state from Charlie to Bob using a procedure that spans the time from $0$ to $T$ as the measurement of the time-non-local observables
\begin{align}\label{eq:varsToMeasure}
&\bar{x}=x_B(T)-x_C(0)&&\t{and}&&\bar{p}=p_B(T)-p_C(0),
\end{align}
and then performing feedback on Bob's oscillator at $t=T$. Alice (the traveling field) is merely the meter in this process.
A measurement of $\bar{x}$ and $\bar{p}$ does not condition a well-defined usual quantum state belonging to $\mathcal{H}=\mathcal{H}_{C,t=0}\otimes\mathcal{H}_{B,t=T}$, the direct product of spaces of ket vectors of Charlie at $t=0$ and Bob at $t=T$, because the operators corresponding to $\bar{x}$ and $\bar{p}$ on $\mathcal{H}$ do not commute. 
Instead, it conditions a two-time state $\Psi$ belonging to $\mathcal{H}'=\mathcal{H}_{C,t=0}^\dagger\otimes\mathcal{H}_{B,t=T}$~\cite{aharonov_multiple-time_2009,aharonov_two-state_2008}, where $\mathcal{H}_{C,t=0}^\dagger$ is the space of Charlie's bra vectors (our two-time states are conjugate compared to the original definition~\cite{aharonov_is_1984,aharonov_two-state_2008}). The operators corresponding to $\bar{x}$ and $\bar{p}$ on $\mathcal{H}'$ do commute. 

To find an explicit expression for $\Psi$, we calculate the product of two time-non-local projectors $\Pi$ enforcing Eqs.~(\ref{eq:varsToMeasure}),
\begin{equation}\label{eq:proj}
\Pi_{x_B(T)-x_C(0)=\bar{x}}\,\Pi_{p_B(T)-p_C(0)=\bar{p}}\propto\Psi\, \Psi^\dagger,
\end{equation}
where
\begin{equation}
\Pi_{x_B(T)-x_C(0)=\bar{x}}=\int\Big(\ket{x+\bar{x}}\bra{x+\bar{x}}\Big)_{B,T}\,\Big(\ket{x}\bra{x}\Big)_{C,0} \, dx,
\end{equation}
and the expression for $\Pi_{p_B(T)-p_C(0)=\bar{p}}$ is analogous. This calculation yields the two-time state
\begin{equation}
\Psi = \int e^{i\bar{p}x}\ket{x+\bar{x}}_{B,t=T}\;\bra{x}_{C,t=0}\, dx,
\end{equation}
which acts as a quantum channel mapping the initial state of Charlie on the final state of Bob (see SI, Sec.~\ref{sec:si:teleport-nl-meas}).
Between two harmonic oscillators, it can be created using only linear interactions, homodyne measurements, and input fields in Gaussian states, while it is able to transfer arbitrary (including non-Gaussian) input states.

\paragraph*{Continuous measurements ---}
The time-non-local measurements of $\bar{x}$ and $\bar{p}$ required to perform teleportation (and, more generally, conditional state transfer) can be implemented via continuous measurements. From this point on, to make the description more symmetric, we label the oscillators 1 and 2 instead of Bob and Charlie, and consider measurements of the sums rather than differences of their $x$ and $p$ quadratures, since their common relative sign is only a matter of convention. We neglect the detection losses and the intrinsic decoherence of the oscillators during the interaction.

The oscillators are described by annihilation operators, $\hat{b}_1$ and $\hat{b}_2$, and have identical frequencies $\Omega$. Each of them linearly interacts with the field, described by annihilation operator $\hat{s}$, via the Hamiltonian
\begin{equation}\label{eq:mainHam}
\hat{H}_\t{int}= \mu(\hat{s}^\dagger \hat{b}+\hat{b}^\dagger \hat{s}) + \nu(\hat{s}^\dagger \hat{b}^\dagger+\hat{s} \hat{b}),
\end{equation}
where $\mu$ and $\nu$ are real, non-negative and, in general, time-dependent. The field satisfies the commutator $[\hat{s}(t),\hat{s}^\dagger(t')]=\delta(t-t')$ and the input-output relation $\hat{s}_{\t{out}}=\hat{s}_{\t{in}}-i(\mu\, \hat{b}+\nu\,\hat{b}^\dagger)$; its input state is vacuum or thermal. 
The interaction is additionally characterized by the measurement rate $\Gamma$, the type $\zeta\in[-1,1]$, interpolating between beamsplitter ($\zeta=1$), position-measurement ($\zeta=0$), and entanglement ($\zeta=-1$) interactions, and the optical damping rate $\gamma$,
\begin{align}\label{eq:rate-defs}
&\Gamma=\frac{(\mu+\nu)^2}{2},&
&\zeta=\frac{\mu-\nu}{\mu+\nu},&
&\gamma=2\zeta\Gamma.
\end{align}
The Hamiltonian~(\ref{eq:mainHam}) describes a range of physical systems including optomechanical cavities~\cite{thomas_entanglement_2020}, gravitational wave detectors, and atomic ensembles~\cite{hammerer_quantum_2010}. In cavity optomechanics, $\zeta=0$ corresponds to the probe laser tuned to the cavity resonance, and $\zeta=1$ ($\zeta=-1$) to the laser red-(blue-)detuned from the resonance by one mechanical frequency in the sideband-resolved regime. In all cases mentioned, the measurement rate $\Gamma$ and the damping $\gamma$ are parametrically controlled by the input optical power. We therefore limit our consideration to time-dependent $\Gamma(t)$ [and $\gamma(t)$] and time-independent $\zeta$. The time dependence of the measurement rates is a crucial part of our analysis and can be seen as a means of matching the temporal field modes that interact with the two oscillators.

Measurements on the oscillators are made via homodyne detection of the output fields. In the sequential configuration (\figref{fig:QM-tasks}d), the demodulated photocurrent constitutes the complex measurement record $m(t)=\hat{p}_\t{out}(t)\,e^{i\Omega t}$, given by
\begin{equation}\label{eq:measRecord}
m(t)=\hat{p}_\t{in}(t)\,e^{i\Omega t}-\sqrt{\Gamma_1(t)}\,\hat{b}_{I,1}(t)-\sqrt{\Gamma_2(t)}\,\hat{b}_{I,2}(t),
\end{equation}
where $\hat{p}_\t{in(out)}(t)=[-i\hat{s}_\t{in(out)}(t)+i\hat{s}_\t{in(out)}^\dagger(t)]/\sqrt{2}$ are the phase quadratures of the input and the output fields, and $\hat{b}_{I,j}=\hat{b}_je^{i\Omega t}$, $j\in\{1,2\}$, are the slowly varying annihilation operators in the interaction picture. 
The continuous record $m(t)$ contains information that can be irrelevant to a given task, and the relevant measurement outcome $\mathcal{M}$ is obtained after applying a filter $f(t)$ as $\mathcal{M}=\int_0^T f(t) m(t) \, dt$. Finding the appropriate $f(t)$ is part of the task.
In the parallel configuration (\figref{fig:QM-tasks}e), measurement records similar to \eqref{eq:measRecord} are obtained and processed analogously (see SI, Sec.~\ref{sec:si:parallel}).

The value $\mathcal{M}$ can correspond to the outcome of a time-non-local measurement.
To see this, we need to express one of the $\hat{b}_{I,j}(t)$ in \eqref{eq:measRecord} via its initial and the other via its final condition by integrating their evolution equations derived from the Hamiltonian~(\ref{eq:mainHam}).
Although we are free to choose either boundary condition when expressing the evolution of each oscillator, for given $\Gamma_{1,2}(t)$ and $\zeta_{1,2}$ at most one choice will allow perfect time-non-local measurements (i.e., with the contribution from the field degrees of freedom vanishing as the measurement strength increases).

\begin{figure}
    \centering
    \includegraphics[width=\columnwidth]{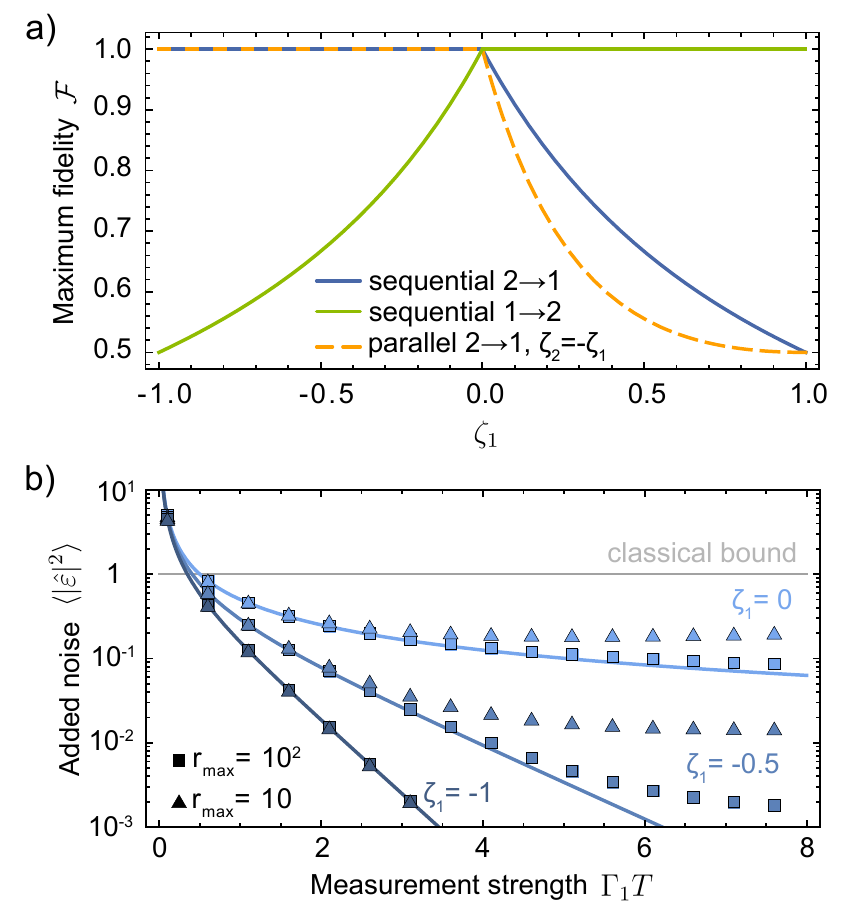}
    \caption{\small{a) The state-transfer fidelities for teleportation [Eqs.~(\ref{eq:telepSeqErr}) and (\ref{eq:telepParErr})] and direct transmission [\eqref{eq:condSeqErr}] for infinite measurement strength $\Gamma_1 T\to\infty$. b) The minimum error for teleportation in the sequential configuration [\eqref{eq:telepSeqErr}] as a function of the measurement strength assuming a vacuum input field. The solid lines represent the idealized case with divergent $\Gamma_2$, while the values indicated by triangles and squares are obtained with $\Gamma_2$ truncated at a finite value $r_\t{max}=\Gamma_{2,\t{max}}/\Gamma_1=10\text{ and }10^2$, respectively (see SI, Sec.~\ref{sec:si:finiteMeasRate}).}}
    \label{fig:bounds}
\end{figure}

\paragraph*{Optimal state transfer by pretrodiction ---}
We can determine to what extent a certain time-non-local observable can be measured with the help of variational analysis. Considering the teleportation of a state from oscillator 2 to 1, we express $\mathcal{M}$ in terms of $\hat{b}_{I,1}(T)$ and $\hat{b}_{I,2}(0)$,
\begin{equation}
\mathcal{M} 
=\hat{\varepsilon}-M_1\, \hat{b}_{I,1}(T)-M_2\, \hat{b}_{I,2}(0),
\end{equation}
where $M_{1,2}$ are the transfer coefficients, and $\hat{\varepsilon}$ is an operator that absorbs all degrees of freedom of the input field, whose variance determines the measurement error. 
Both oscillators can have arbitrary initial states (not chosen from a restricted family), and therefore we impose the constraint $M_1=M_2=1$. Under this constraint, $\mathcal{M}$ is related to the pretrodiction observables equivalent to those in Eqs.~(\ref{eq:varsToMeasure}) via $\mathcal{M}=\hat{\varepsilon}-(\bar{x}+i\bar{p})\sqrt{2}$, where the relevant positions and momenta are the rotating-frame quadratures of each oscillator,
\begin{align}
&\hat{x}=(\hat{b}_I+\hat{b}_I^\dagger)/\sqrt{2},
&\hat{p}=(\hat{b}_I-\hat{b}_I^\dagger)/(i\sqrt{2}).
\end{align}
The measurement error for either of $\bar{x}$ and $\bar{p}$ is given by $\langle\abs{\hat{\varepsilon}}^2\rangle\equiv\langle(\hat{\varepsilon}\hat{\varepsilon}^\dagger+\hat{\varepsilon}^\dagger \hat{\varepsilon})/2\rangle$, where the averaging is over the input state of the field. The state-transfer fidelity~\cite{hammerer_quantum_2010} is found from this error as 
$\mathcal{F}\equiv 1/(1+\langle\abs{\hat{\varepsilon}}^2\rangle)$. 
To find the maximum fidelity, we analytically perform the nonlinear, constrained variational minimization
\begin{align}
&\langle \abs{\hat{\varepsilon}}^2 \rangle\to \t{min},
&M_1=M_2=1,
\end{align}
over $f(t)$ and $\Gamma_2(t)$ (see SI, \secref{sec:si:SeqTel}). The measurement rate of only one of the oscillators needs to change in time; we choose $\Gamma_2$ to be time-dependent, and $\Gamma_1$ to be constant.

In the sequential configuration (\figref{fig:QM-tasks}d) with vacuum input fields, the minimum error for teleportation is
\begin{equation}\label{eq:telepSeqErr}
\langle\abs{\hat{\varepsilon}}^2\rangle_{\t{seq},2\to 1}=\frac{\zeta_1}{1-\exp(-\gamma_1 T)}.
\end{equation}
The error depends on the interaction type of the first oscillator $\zeta_1$, but not on that of the second, $\zeta_2$. 
Whenever $\zeta_1<0$, i.e., the interaction of first oscillator is entanglement-dominated, the teleportation fidelity approaches one as the measurement strength is increased, $\Gamma_1 T\to\infty$.
In the parallel configuration (\figref{fig:QM-tasks}e), the additional condition $\zeta_2=-\zeta_1$ must be satisfied for the fidelity to approach one as $\Gamma_1 T\to\infty$, in which case the minimum error is
\begin{equation}\label{eq:telepParErr}
\langle\abs{\hat{\varepsilon}}^2\rangle_{\t{par},2\to 1}=\frac{2\zeta_1/(1+\zeta_1^2)}{1-\exp(-\gamma_1 T)}.
\end{equation}
The optimum filter function $f(t)$ is in both cases
\begin{equation}\label{eq:fTel}
f(t)=\sqrt{\Gamma_1}\frac{ (\zeta_1+\zeta_2)e^{\gamma_1 t/2}+(\zeta_1-\zeta_2)e^{-\gamma_1 t/2}}{2\sinh(\gamma_1 T/2)},
\end{equation}
and the required time-dependent $\Gamma_2$ is
\begin{equation}\label{eq:GammaOptTel}
\frac{\Gamma_2(t)}{\Gamma_1}=\frac{\sinh^2(\gamma_1 t/2)}{\sinh^2(\gamma_1 T/2)-\sinh^2(\gamma_1 t/2)}.
\end{equation}
While $\Gamma_2(t)$ diverges as $t\to T$, it can be truncated at some finite value. The effect of this is a fidelity reduction that depends on the truncation point and can be made negligible (see \figref{fig:bounds}b and SI, Sec.~\ref{sec:si:finiteMeasRate}).

The conditional direct transmission of the state from oscillator 1 to 2 in the sequential configuration (\figref{fig:QM-tasks}d) is treated similarly to the teleportation; it does not have a counterpart in the parallel configuration. The minimum error,
\begin{equation}\label{eq:condSeqErr}
\langle\abs{\hat{\varepsilon}}^2\rangle_{\t{seq},1\to 2}=\frac{\zeta_1}{\exp(\gamma_1 T)-1},
\end{equation}
is independent of $\zeta_2$ and realized by the filter function $f(t)$ and the measurement rate $\Gamma_2(t)$ that are obtained from Eqs.~(\ref{eq:fTel}) and (\ref{eq:GammaOptTel}) by the replacement $t\to t-T$ (see SI, \secref{sec:si:SeqDir}).

When the input field is in a thermal state with a non-zero population $n_\t{in}$, as in the case of microwave transmission lines~\cite{vermersch_quantum_2017,magnard_microwave_2020}, the errors in Eqs.~(\ref{eq:telepSeqErr}), (\ref{eq:telepParErr}), and (\ref{eq:condSeqErr}) are multiplied by ($2n_\t{in}$+1). 
For concreteness, we will keep assuming $n_\t{in}=0$, as is typical of optical fields, when presenting the results.

The performance of our state-transfer protocols is summarized in \figref{fig:bounds}. Panel a) shows the teleportation and direct transfer fidelities for $\Gamma_1 T\to \infty$. 
The parameter regions where unit fidelity can be reached for teleportation, $\zeta_1\le0$, and direct transfer, $\zeta_1\ge0$, intersect at $\zeta_1=0$. However, this does not permit perfect \emph{exchange} of the initial states between the oscillators, because different directions of the state transfer require different time dependencies of the measurement rate.
E.g., if an arbitrary initial state of oscillator 2 is perfectly teleported to oscillator 1 by measuring $\hat{b}_{I,1}(T) + \hat{b}_{I,2}(0)$, the error for the measurement of $\hat{b}_{I,1}(0)+\hat{b}_{I,2}(T)$ is infinite. 
Figure~\ref{fig:bounds}b shows the error for finite $\Gamma_1 T$ for teleportation in the sequential configuration. In the regime dominated by optical anti-damping of oscillator 1, $\zeta_1\gtrsim -1$, the teleportation error decreases exponentially $\langle\abs{\hat{\varepsilon}}^2\rangle\approx e^{-\Gamma_1 T}$, whereas in the position-measurement regime, $\zeta_1\approx0$, the error decreases inversely proportional to the measurement strength $\langle\abs{\hat{\varepsilon}}^2\rangle\approx 1/(2\Gamma_1 T)$.

\paragraph*{How conditional is the state transfer? ---}

\begin{figure}
    \centering
    \includegraphics[width=\columnwidth]{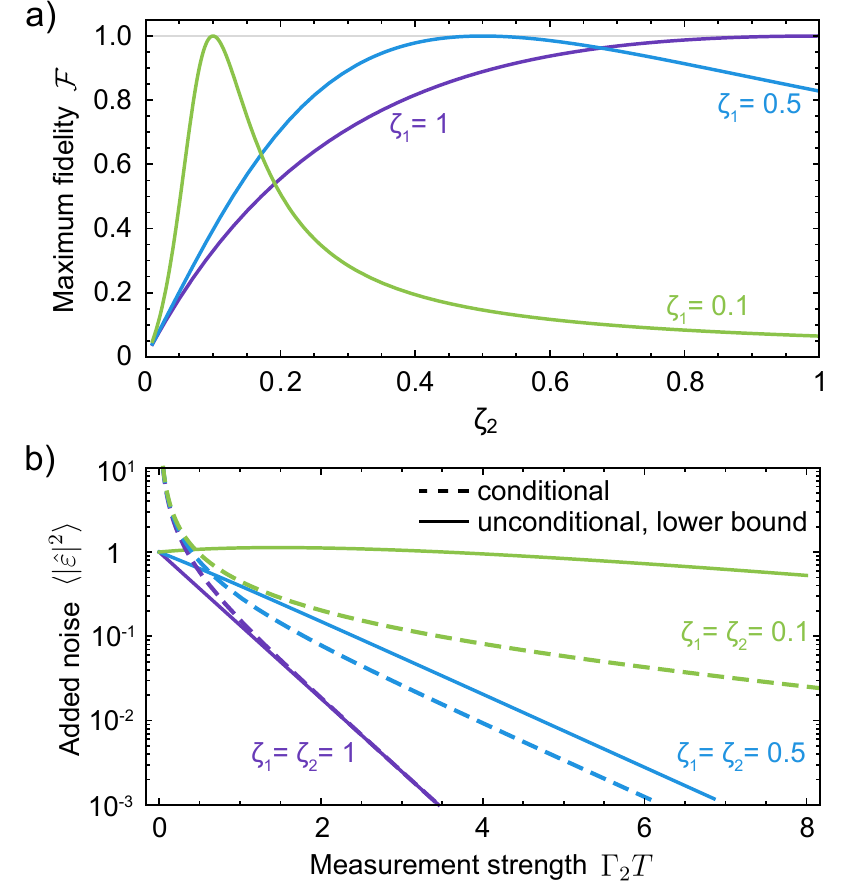}
    \caption{\small{a) The state-transfer fidelities of unconditional direct transmission for different $\zeta_{1,2}$. b) The minimum added noise during conditional and unconditional direct state transfer, respectively, as function of the measurement strength for $\zeta_1=\zeta_2$. 
    }}
    \label{fig:ballistic}
\end{figure}

The variances of the outcomes $\bar{x}$ and $\bar{p}$ of perfect pretrodiction measurements with $\langle\abs{\hat{\varepsilon}}^2\rangle\to 0$ determine the state-transfer error if these outcomes were to be discarded. In this case, the \emph{unconditional} state-transfer fidelity~\cite{hammerer_quantum_2010} is given by $\mathcal{F}_\t{uc}=1\left/\left(1+\langle|\mathcal{M}|^2 \rangle\right)\right.$. 
This fidelity depends on the interaction setting and the input state of the traveling field. 
In the teleportation settings, $\langle|\mathcal{M}|^2 \rangle\to\infty$ and $\mathcal{F}_\t{uc}=0$, consistent with the causality principle. 
In the direct transfer setting ($1\to2$, \figref{fig:QM-tasks}d), $\mathcal{F}_\t{uc}$ is generally non-zero and even reaches one, meaning that the unconditional state transfer can be perfect. 
To see this, in \figref{fig:ballistic}a we compare the unconditional state-transfer fidelities in the sequential configuration for different interaction types $\zeta_{1,2}$, infinite measurement strength, and optimum rates $\Gamma_{1,2}$ minimizing the pretrodiction measurement error (this choice generally does not optimize the unconditional performance).
We let the measurement rate of the second oscillator $\Gamma_2$ be constant, and allow $\Gamma_1(t)$ to change in time (see SI, Sec.~\ref{sec:si:uncond}); in this case the measurement strength is given by $\Gamma_2 T$. 
We find that $\mathcal{F}_\t{uc}$ can be above zero when $\zeta_{1,2}>0$, and that $\mathcal{F}_\t{uc}=1$ when $\zeta_1=\zeta_2>0$.
In the specific case $\zeta_1=\zeta_2=1$, which describes, e.g., catching a state leaking from one electromagnetic cavity with another, our expression for the time-dependent $\Gamma_1(t)$ converges as $\Gamma_2 T\to \infty$ to the solution obtained in Ref.~\cite{jahne_high-fidelity_2007} using a different fidelity measure (see SI, Sec.~\ref{sec:si:comp-direct}).

Even though a perfect state transfer may be unconditional as $\Gamma_2 T\to \infty$, at finite measurement strengths discarding the outcomes $\bar{x}$ and $\bar{p}$ introduces extra error. This is illustrated in \figref{fig:ballistic}b, where we compare the conditional and unconditional state-transfer errors for $\zeta_1=\zeta_2>0$, the case in which both go to zero as $\Gamma_2 T\to\infty$. The figure shows that, as expected, the unconditional errors are always higher than the conditional errors at large measurement strengths. 
At small measurement strengths, our method of comparison breaks down. While the unconditional errors cannot be evaluated for arbitrary initial states, the fact that the initial-state contribution to the error decays over time allows us to lower-bound it by assuming that the initial states of both oscillators are vacuum. Such prior knowledge is absent in the evaluation of the conditional error, whence it can exceed the unconditional value at small measurement strengths (see SI, Sec.~\ref{sec:si:uncond}).

\paragraph*{Conclusions and outlook ---}
We introduced prediction-retrodiction measurements as a new primitive in quantum measurement theory and showed its application to the problem of state transfer between localized oscillators. A similar analysis may yield new insights into discrete-variable protocols. Elements of such an approach can be found in Ref.~\cite{greplova_quantum_2016}, where retrodiction was applied in the Bell-measurement step of teleportation between two qubits coupled to the same cavity. To extend this to a pretrodiction analysis would require incorporating the prediction component constituted by the entanglement step.

In the settings considered, the state transfer is only possible one way, from oscillator 1 to 2, or 2 to 1. However, Vaidman has proposed a two-way teleportation scheme based on ``crossed'' time-non-local measurements that accomplishes exchange of initial states~\cite{vaidman_teleportation_1994}. The pretrodiction framework may help to find a realization of this scheme via continuous measurements.

The result that perfect conditional and unconditional state transfer between oscillators can be realized with a wide range of linear couplings between oscillators and fields, i.e., not only with pure beamsplitter or two-mode-squeezing interaction, can find applications in emergent optical and microwave quantum information processing. In particular, the resilience of our protocols to thermal noise in the input field is a useful trait for microwave quantum links~\cite{vermersch_quantum_2017,magnard_microwave_2020}. 

\paragraph*{Acknowledgements ---}
The authors acknowledge fruitful discussions with Eugene Polzik and Micha\l{} Parniak.
This work was supported by the European Research Council (ERC) under the Horizon 2020 (grant agreement No 787520) and by VILLUM FONDEN under a Villum Investigator Grant no.\ 25880.
S.\,A.\,F.\ acknowledges funding from the European Union’s Horizon 2020 research program under the Marie Sklodowska-Curie grant agreement No.\ 847523 ``INTERACTIONS".

\bibliography{references}

\appendix

\setcounter{equation}{0}
\setcounter{figure}{0}
\renewcommand{\theequation}{S\arabic{equation}}
\renewcommand{\thefigure}{S\arabic{figure}}

\onecolumngrid

\section*{Supplementary Information}

\subsection{Teleportation as a result of time-non-local measurements}\label{sec:si:teleport-nl-meas}

To establish equivalence between teleportation and the result of a time-non-local measurement, we first show how the teleportation protocol introduced by Bennett, Brassard, Crepeau, Jozsa, Peres, and Wootters (BBCJPW)~\cite{bennett_teleporting_1993} and generalized to continuous-variable systems by Vaidman~\cite{vaidman_teleportation_1994} and by Braunstein and Kimble~\cite{braunstein_teleportation_1998} leads to the creation of a certain multiple-time state~\cite{aharonov_two-state_2008}, which we call the ``transfer state", and then show how this transfer state can be produced via time-non-local measurements.

\subsubsection{The multiple-time state created in BBCJPW teleportation}
At the outset of the BBCJPW protocol, at the time $t_1$, two oscillators, possessed respectively by Alice and Bob, are in the EPR state with $x_A-x_B=0$ and $p_A+p_B=0$,
described by the unnormalized expression
\begin{equation}
\ket{\psi_\t{EPR}}_{AB,t_1}=\int dx\, \ket{x}_{A,t_1}\ket{x}_{B,t_1}.
\end{equation}
In combination with the initial state of Charlie's oscillator $\ket{\psi_\t{in}}_{C,t_1}$, the input state to be teleported, the (unnormalized) joint state of the entire system is
\begin{equation}\label{eq:si:psi-ABC-t1}
\ket{\psi}_{ABC,t_1}=\ket{\psi_\t{EPR}}_{AB,t_1}\ket{\psi_\t{in}}_{C,t_1}.
\end{equation}
From $t_1$ to the final time $t_2$ all subsystems evolve independently according to their unitary free-evolution operators $\hat{U}_j(t_2,t_1)$ with $j\in\{A,B,C\}$; in the original formulation of the BBCJPW protocol, the evolution operators are identities. At time $t_2$, the states of Alice and Charlie are projected by a Bell measurement on the state
\begin{equation}
\ket{\psi_\t{Bell}}_{AC,t_2}=\int dx\, e^{-i\bar{p} x}\ket{x+\bar{x}}_{A,t_2}\ket{x}_{C,t_2},
\end{equation}
characterized by the measurement outcomes $\bar{x}=x_A-x_C$ and $\bar{p}=-(p_A+p_C)$.
Applying the corresponding projector $\ket{\psi_\t{Bell}}_{AC,t_2}\bra{\psi_\t{Bell}}_{AC,t_2}$ to the time-evolved initial state $\ket{\psi}_{ABC,t_2}=\hat{U}(t_2,t_1)\ket{\psi}_{ABC,t_1}$, where $\hat{U}(t_2,t_1)=\prod_j\hat{U}_j(t_2,t_1)$, we arrive at the final conditional state of the entire system
\begin{equation}\label{eq:si:psi-ABC-f}
    \ket{\psi}_{ABC,\t{f}} = \ket{\psi_\t{Bell}}_{AC,t_2}\underbrace{\bra{\psi_\t{Bell}}_{AC,t_2} \hat{U}(t_2,t_1)\ket{\psi_\t{EPR}}_{AB,t_1}}_{\Psi\equiv}\ket{\psi_\t{in}}_{C,t_1},
\end{equation}
completing the teleportation protocol (up to a coherent displacement of Bob's state). A similar wave-function exposition of the teleportation protocol including a finite degree of EPR entanglement can be found in Ref.~\cite{vaidman_another_2006}.

The protocol presented above works for an arbitrary input state $\ket{\psi_\t{in}}_{C,t_1}$, initially in a product state with respect to Alice and Bob, and, at the final time, leaves Bob's system in a product state with respect to Alice and Charlie, which have served their purpose in the protocol (and are now maximally entangled). 
The transition to the multiple-time formalism~\cite{aharonov_two-state_2008} consists in eliminating the meter degree of freedom, Alice, along with Charlie's initial and final states. The object that accomplishes this, $\Psi$ in \eqref{eq:si:psi-ABC-f}, is the multiple-time state conditioned by the measurement, which involves only the systems of Charlie and Bob,
\begin{align}
\Psi&=\bra{\psi_\t{Bell}}_{AC,t_2} \hat{U}(t_2,t_1)\ket{\psi_\t{EPR}}_{AB,t_1} \nonumber\\
&=\int dxdx_1\, e^{i\bar{p} x_1}\bra{x_1+\bar{x}}_{A,t_2}\bra{x_1}_{C,t_2}\hat{U}(t_2,t_1) \ket{x}_{A,t_1}\ket{x}_{B,t_1}=\int dx\, e^{i\bar{p} x} \ket{x+\bar{x}}_{B,t_2}\bra{x}_{C,t_1}.
\end{align}
Viewed as an operator (the duality between multiple-time states and operators is discussed in Ref.~\cite{aharonov_multiple-time_2009}), $\Psi$ can be applied to a usual input quantum state of Charlie's oscillator to conditionally transfer it to Bob's. An application of feedback to Bob's oscillator makes the transfer unconditional by removing the displacements $\bar{x}$ and $\bar{p}$.

\subsubsection{The creation of transfer states by time-non-local measurements}

The two-time states that perform conditional state transfer, including teleportation as seen in the preceding subsection,
\begin{equation}
\Psi=\int dx e^{i\bar{p} x}\ket{x+\bar{x}}_{B,t_2}\,\bra{x}_{C,t_1},
\end{equation}
are projected upon by the time-non-local measurements of $x_B(t_2)-x_C(t_1)$ and $p_B(t_2)-p_C(t_1)$ with the outcomes $\bar{x}$ and $\bar{p}$, respectively. We will now demonstrate this by explicitly comparing the (unnormalized) projector upon $\Psi$,
\begin{equation}\label{eq:si:Pi_Psi}
\Pi_{\Psi}\equiv\Psi\, \Psi^\dagger
=\int dx_1dx_2\, e^{i\bar{p} (x_1-x_2)}\Big(\ket{x_1+\bar{x}}\bra{x_2+\bar{x}}\Big)_{B,t_2}\Big(\ket{x_2}\bra{x_1}\Big)_{C,t_1},
\end{equation}
and the product of the projectors $\Pi_{x_B(t_2)-x_C(t_1)=\bar{x}}$ and $\Pi_{p_B(t_2)-p_C(t_1)=\bar{p}}$ collapsing the state upon the measurements of the position and momentum differences, where
\begin{align}
\Pi_{x_B(t_2)-x_C(t_1)=\bar{x}}&=\int dx\,\Big(\ket{x+\bar{x}}\bra{x+\bar{x}}\Big)_{B,t_2}\,\Big(\ket{x}\bra{x}\Big)_{C,t_1},\\
\Pi_{p_B(t_2)-p_C(t_1)=\bar{p}}&=\int dp\,\Big(\ket{p+\bar{p}}\bra{p+\bar{p}}\Big)_{B,t_2}\,\Big(\ket{p}\bra{p}\Big)_{C,t_1}.
\end{align}
The product of $\Pi_{x_B(t_2)-x_C(t_1)=\bar{x}}$ and $\Pi_{p_B(t_2)-p_C(t_1)=\bar{p}}$ is found as
\begin{equation}\label{eq:si:PipProd}
\Pi_{x_B(t_2)-x_C(t_1)=\bar{x}}\; \Pi_{p_B(t_2)-p_C(t_1)=\bar{p}}
=\int dxdp\,\Big(\ket{x+\bar{x}}\langle x+\bar{x}\vert p+\bar{p}\rangle\bra{p+\bar{p}}\Big)_{B,t_2}\,\Big(\ket{p}\langle p\vert x\rangle\bra{x}\Big)_{C,t_1}.
\end{equation}
Note that in \eqref{eq:si:PipProd} the components of the projectors acting at different times are combined from different sides, because the multiple-time states that they apply to live in the product of the Hilbert space of Bob and the conjugate Hilbert space of Charlie $\mathcal{H}_{B,t_2}\otimes\mathcal{H}_{C,t_1}^\dagger$~\cite{aharonov_multiple-time_2009}.
Using the relationship between position and momentum eigenstates $\langle x|p \rangle =e^{ipx}/\sqrt{2\pi}$, \eqref{eq:si:PipProd} can be further rewritten to be indeed proportional to $\Pi_{\Psi}$ [\eqref{eq:si:Pi_Psi}],
\begin{equation}
\Pi_{x_B(t_2)-x_C(t_1)=\bar{x}}\; \Pi_{p_B(t_2)-p_C(t_1)=\bar{p}}
=(2\pi)^{-1}\,\Pi_\Psi.
\end{equation}
The proportionality constant is insignificant because the states that we consider were not normalized from the beginning.

\subsection{The Bogoliubov modes of the traveling field and equation of motion for the localized oscillators}

To solve the dynamics of oscillators coupled to propagating fields by general linear Hamiltonians, we introduce the (unnormalized) Bogoliubov operators of the input field $\hat{u}^\zeta_\t{in}$,
\begin{equation}\label{eq:si:BogDef}
\hat{u}^\zeta_\t{in}(t)=\frac{1}{\sqrt{2}}\left((1+\zeta) \hat{s}_\t{in}(t)+(1-\zeta) \hat{s}_\t{in}^\dagger(t)\right)e^{i\Omega t}.
\end{equation}
Their two-time commutators are
\begin{equation}\label{eq:si:BogComm}
\left[\hat{u}^\zeta_\t{in}(t),\left(\hat{u}^{\zeta'}_\t{in}(t')\right)^\dagger\right]=(\zeta+\zeta')\delta(t-t'),
\end{equation}
and their symmetrized correlation functions are
\begin{equation}\label{eq:si:BogCorr}
\avg{\left(\hat{u}^\zeta_\t{in}(t)\right)^\dagger \hat{u}^{\zeta'}_\t{in}(t')+\hat{u}^{\zeta'}_\t{in}(t') \left(\hat{u}^\zeta_\t{in}(t)\right)^\dagger}=
\left(1+\zeta\zeta'\right)\left(2n_\t{in}+1\right)\,\delta(t-t').
\end{equation}
Here $n_\t{in}$ is the thermal equilibrium occupation of the traveling field. For brevity, we will present the expressions for the added noise in the state transfer protocols for the case of $n_\t{in}=0$ (typical of optical implementations), while the general expressions can always be recovered by multiplying the variance of the errors by $\left(2n_\t{in}+1\right)$ (relevant for, e.g., microwave implementations).

For arbitrary $\zeta$, the measurement record $m(t)$ obtained by demodulating the photocurrent of a homodyne detector tuned to detect the phase quadrature of the light, can be expressed via the output Bogoliubov modes [defined analogously to their input counterparts~(\ref{eq:si:BogDef})] as
\begin{equation}\label{eq:si:m-general}
m(t)=\frac{1}{\sqrt{2}}\left(-i\hat{s}_{\t{out}}(t)+i\hat{s}_{\t{out}}^\dagger(t)\right)e^{i\Omega t}=\frac{1}{2\zeta}\left(-i\hat{u}^{\zeta}_\t{out}(t)+i\hat{u}^{-\zeta}_\t{out}(t)\right).
\end{equation}

We now turn to the evolution equation governing an individual localized oscillator described by the annihilation operator $\hat{b}$. The free oscillation at the frequency $\Omega$ combined with the damping $\gamma$ and fluctuations $\propto \hat{s}_\t{in}$ and $\hat{s}_\t{in}^\dagger$ induced by the oscillator-field interaction [\eqref{eq:mainHam} in the main text] result in the lab-frame Heisenberg-Langevin equation of motion
\begin{equation}\label{eq:si:EOM-lab}
\frac{d}{dt}\hat{b}(t)=-\left(i\Omega+\frac{\gamma(t)}{2}\right)\hat{b}(t)-i\left(\mu(t)\hat{s}_{\t{in}}(t) +\nu(t)\hat{s}_{\t{in}}^\dagger(t)\right). 
\end{equation}
The input-output relation for the field is 
\begin{equation}\label{eq:si:IO-rel}
\hat{s}_{\t{out}}(t)=\hat{s}_{\t{in}}(t)-i\left(\mu(t)\, \hat{b}(t)+\nu(t)\,\hat{b}^\dagger(t)\right).
\end{equation}
Introducing the slowly varying annihilation operator $\hat{b}_I(t)=e^{i\Omega t}\hat{b}(t)$, as in the main text, the oscillator evolution in the interaction frame can be described in terms of the above Bogoliubov modes (\ref{eq:si:BogDef}) of the input light field,
\begin{equation}\label{eq:si:EOM-rot}
    \frac{d}{dt}\hat{b}_{I}(t)=-\frac{\gamma(t)}{2}\hat{b}_{I}(t)-i\sqrt{\Gamma(t)}\,\hat{u}^{\zeta}_{\t{in}}(t),
\end{equation}
which also uses the parametrization of $\mu=(1+\zeta)\sqrt{\Gamma/2}$ and $\nu=(1-\zeta)\sqrt{\Gamma/2}$ in terms of the measurement rate $\Gamma$ and the interaction type $\zeta$ as defined in Eqs.~(\ref{eq:rate-defs}) in the main text.

\subsection{The sequential case: teleportation}\label{sec:si:SeqTel}

In the sequential configuration [Fig.~\ref{fig:QM-tasks}d in the main text], one input field interacts sequentially with two oscillator modes, $\hat{b}_{1}$ and $\hat{b}_{2}$ (in this order), i.e., the input field for oscillator 2 is the output field from oscillator 1,
\begin{equation}\label{eq:si:seq-link}
\hat{s}^{(2)}_\t{in}(t)=\hat{s}^{(1)}_\t{out}(t),
\end{equation}
ignoring the time delay due to propagation between the two systems. 
Using this and the input-output relation~(\ref{eq:si:IO-rel}),
the oscillator evolution equations~(\ref{eq:si:EOM-rot}) can be stated in terms of Bogoliubov modes $\hat{u}^{\zeta}_{\t{in}}$ of the input field of oscillator 1 $\hat{s}_\t{in}\equiv \hat{s}^{(1)}_\t{in}$,
\begin{align}
\frac{d}{dt}\hat{b}_{I,1}(t)&=-\frac{\gamma_1(t)}{2}\hat{b}_{I,1}(t)-i\sqrt{\Gamma_1(t)}\,\hat{u}^{\zeta_1}_{\t{in}}(t),\label{eq:si:b1}\\
\frac{d}{dt}\hat{b}_{I,2}(t)&=-\frac{\gamma_2(t)}{2}\hat{b}_{I,2}(t)-\sqrt{\Gamma_1(t)\Gamma_2(t)}(\zeta_1+\zeta_2)\hat{b}_{I,1}(t)
-i\sqrt{\Gamma_2(t)}\hat{u}^{\zeta_2}_{\t{in}}(t).\label{eq:si:b2}
\end{align}
Above we have assumed the rotating-wave approximation, valid whenever $|\Omega|\gg \Gamma_j(2n_\t{in}+1)$.
The measurement rates $\Gamma_j$ and the optical damping rates $\gamma_j$ are proportional to each other $\gamma_j(t)=2\zeta_j\Gamma_j(t)$ and generally functions of time [Eqs.~(\ref{eq:rate-defs})]; the $\zeta_j$, characterizing the type of interaction, are assumed constant.

There is one complex measurement record $m(t)$ produced as a result of the homodyne detection of the phase quadrature of the output light,
\begin{equation}\label{eq:si:meas}
m(t)=\frac{1}{2\zeta_1}\left(-i\hat{u}^{\zeta_1}_\t{in}(t)+i\hat{u}^{-\zeta_1}_\t{in}(t)\right)
-\sqrt{\Gamma_1(t)}\hat{b}_{I,1}(t)-\sqrt{\Gamma_2(t)}\hat{b}_{I,2}(t),
\end{equation}
as follows from Eq.~(\ref{eq:si:m-general}), with the particular choice $\zeta=\zeta_1$, combined with Eqs.~(\ref{eq:si:seq-link}) and (\ref{eq:si:IO-rel}) along with the definitions~(\ref{eq:rate-defs}) from the main text.
It is convenient to rescale the regular time $t$ to a dimensionless ``interaction'' time variable proportional to the mean accumulated number of drive photons that have impinged on the system up until time $t$,
\begin{equation}\label{eq:si:tau-def}
\tau_j(t)=\int_{0}^{t} \gamma_j(t')dt'.
\end{equation}
Informed by the contribution of Bob's system to the pretrodiction observable, the equation of motion~(\ref{eq:si:b1}) for the first oscillator is solved by backward-evolving the final annihilation operator $\hat{b}_{I,1}(T)$,
\begin{equation}\label{eq:si:b1-sol-seq}
\hat{b}_{I,1}(t)=e^{(\tau_1(T)-\tau_1(t))/2}\hat{b}_{I,1}(T)+
ie^{-\tau_1(t)/2}\int_{t}^{T}e^{\tau_1(t')/2}\sqrt{\Gamma_1(t')}\,\hat{u}^{\zeta_1}_\t{in}(t')dt'.
\end{equation}
Dealing with the second oscillator, we first use \eqref{eq:si:meas} for the measurement record to rewrite \eqref{eq:si:b2} in a form that does not explicitly contain the coupling term proportional to $\hat{b}_{I,1}$,
\begin{equation}\label{eq:si:b2-m}
\frac{d}{dt}\hat{b}_{I,2}(t)=-\frac{\tilde{\gamma}_2(t)}{2}\hat{b}_{I,2}(t)
-i\sqrt{\Gamma_2(t)}\hat{u}^{-\zeta_1}_\t{in}(t)+\sqrt{\Gamma_2(t)}(\zeta_1+\zeta_2) m(t),
\end{equation}
where, for oscillator 2, we have introduced the renormalized damping $\tilde{\gamma}_2$ and interaction time $\tilde{\tau}_2$,
\begin{align}\label{eq:si:tilde2-defs}
&\tilde{\gamma}_2(t)=-\frac{\zeta_1}{\zeta_2} \gamma_2(t),
&\tilde{\tau}_2(t)=-\frac{\zeta_1}{\zeta_2} \tau_2(t).
\end{align}
In this representation, the optical noise driving oscillator 2 is the sum of the operator $\hat{u}^{-\zeta_1}_\t{in}$, which commutes with the input noise driving oscillator 1 [see \eqref{eq:si:BogComm}], and a classical force proportional to the measurement outcome $m(t)$. Informed by the contribution of (Charlie's) oscillator 2 to the pretrodiction observable, the solution at time $t$ is obtained by forward-evolving the initial-time operator $\hat{b}_{I,2}(0)$ using \eqref{eq:si:b2-m},
\begin{equation}\label{eq:si:b2-sol-seq}
\hat{b}_{I,2}(t)=e^{-\tilde{\tau}_2(t)/2}\hat{b}_{I,2}(0)-
ie^{-\tilde{\tau}_2(t)/2}\int_{0}^{t}e^{\tilde{\tau}_2(t')/2}\sqrt{\Gamma_2(t')}\,\hat{u}^{-\zeta_1}_\t{in}(t')dt'+\phi_m(t),
\end{equation}
where the term 
\begin{equation}
\phi_m(t)=(\zeta_1+\zeta_2)e^{-\tilde{\tau}_2(t)/2}
\int_{0}^{t}e^{\tilde{\tau}_2(t')/2}\sqrt{\Gamma_2(t')}\,m(t')\,dt'
\end{equation}
is a deterministic displacement after the measurement record is obtained. 
Inserting \eqref{eq:si:b2-sol-seq} in \eqref{eq:si:meas}, we see that the $\phi_m(t)$ contribution can be absorbed in a renormalized measurement record 
\begin{equation}\label{eq:si:measRecRenorm}
\tilde{m}(t)=m(t)+\sqrt{\Gamma_2(t)}\,\phi_m(t).
\end{equation}
Henceforth, we analyze the time-non-local measurements required for the state transfer by integrating the renormalized measurement record $\tilde{m}(t)$ 
with a filter function $\tilde{f}(t)$, which yields
\begin{equation}\label{eq:si:mathcalM-seq}
\mathcal{M}=\int_{0}^{T}\tilde{f}(t)\tilde{m}(t)dt=
\hat{\varepsilon} - M_1 \hat{b}_{I,1}(T) - M_2 \hat{b}_{I,2}(0).
\end{equation}
The measurement error is set by the (zero-mean) operator $\hat{\varepsilon}$ that only contains optical degrees of freedom. The explicit expressions that we present below for the elements of \eqref{eq:si:mathcalM-seq} follow from inserting Eqs.~(\ref{eq:si:b1-sol-seq}) and (\ref{eq:si:b2-sol-seq}) into \eqref{eq:si:meas}, and then transitioning to the renormalized measurement record $\tilde{m}(t)$ using \eqref{eq:si:measRecRenorm}. From these steps, two central quantities emerge, namely the partial (time-dependent) transfer coefficients
\begin{subequations}\label{eq:si:M-t_seq}
\begin{align}
M_1(t)&=\int_{0}^{t}\tilde{f}(t')\sqrt{\Gamma_1(t')}\,e^{[\tau_1(t)-\tau_1(t')]/2}dt',\\
M_2(t)&=\int_{t}^{T}\tilde{f}(t')\sqrt{\Gamma_2(t')}\,e^{[\tilde{\tau}_2(t)-\tilde{\tau}_2(t')]/2}dt'.
\end{align}
\end{subequations}
The final (time-independent) transfer coefficients $M_1$ and $M_2$ appearing in \eqref{eq:si:mathcalM-seq} are given by appropriate end-point values of the time-dependent $M_j(t)$ [Eqs.~(\ref{eq:si:M-t_seq})],
\begin{align}\label{eq:si:M-final-SerTel}
&M_1\equiv M_1(T),&M_2\equiv M_2(0).
\end{align}
If $M_1$ and $M_2$ are equal and normalized to one ($M_1=M_2=1$), 
then the results of the measurement of the pretrodiction observables $\bar{x}$ and $\bar{p}$ are extracted from $\mathcal{M}$~(\ref{eq:si:mathcalM-seq}) as
\begin{align}
&\bar{x}\equiv-\sqrt{2}\re{\mathcal{M}}=\hat{x}_{1}(T)+\hat{x}_{2}(0)-\hat{\varepsilon}_x, &\bar{p}\equiv-\sqrt{2}\im{\mathcal{M}}=\hat{p}_{1}(T)+\hat{p}_{2}(0)-\hat{\varepsilon}_p, 
\end{align}
and their respective errors are given by the quadrature operators $\hat{\varepsilon}_x=(\hat{\varepsilon}+\hat{\varepsilon}^\dagger)/\sqrt{2}$ and $\hat{\varepsilon}_p=(\hat{\varepsilon}-\hat{\varepsilon}^\dagger)/(\sqrt{2}i)$. 
The measurement enables the transfer of an arbitrary state from oscillator 2 to oscillator 1, with a fidelity determined by the variances of $\hat{\varepsilon}_{x,p}$, by displacing the quadratures of oscillator 1 by $-\bar{x}$ and $-\bar{p}$, respectively. Within the rotating-wave approximation, assumed throughout this work, the error quadrature operators are uncorrelated and of equal variance. 
We will be concerned, therefore, with minimizing the variance of the single-quadrature measurement error 
\begin{equation}\label{eq:si:epsilon-var-def}
\langle\abs{\hat{\varepsilon}}^2\rangle\equiv\avg{\hat{\varepsilon}_x^2}=\avg{\hat{\varepsilon}_p^2}=\frac{1}{2}\avg{\hat{\varepsilon}\hat{\varepsilon}^\dagger + \hat{\varepsilon}^\dagger\hat{\varepsilon}}
\end{equation}
over all possible filter functions $\tilde{f}(t)$ and time-dependent measurement rates $\Gamma_{j}(t)$ under the constraint
\begin{equation}\label{eq:si:M-constraint_seq}
M_1=M_2\equiv M=1.
\end{equation}

The error operator $\hat{\varepsilon}$ in \eqref{eq:si:mathcalM-seq} can, via a change of integration order, be expressed in terms of the input fields as 
\begin{equation}\label{eq:si:optError}
\hat{\varepsilon}=-i\int_{0}^{T}B_1(t)\hat{u}^{\zeta_1}_\t{in}(t)dt
+i\int_{0}^{T}B_2(t)\hat{u}^{-\zeta_1}_\t{in}(t)dt,
\end{equation}
where the temporal noise modes $B_j(t)$ are given in terms of $M_j(t)$ [Eqs.~(\ref{eq:si:M-t_seq})] as
\begin{subequations}\label{eq:si:BSerTel}
\begin{align}
B_1(t)&=\frac{\tilde{f}(t)}{2\zeta_1}+\sqrt{\Gamma_1(t)}M_1(t),\label{eq:si:B1SerTel}\\
B_2(t)&=\frac{\tilde{f}(t)}{2\zeta_1}+\sqrt{\Gamma_2(t)}M_2(t).\label{eq:si:B2SerTel}
\end{align}
\end{subequations}
The variance~(\ref{eq:si:epsilon-var-def}) of  $\hat{\varepsilon}$ can be calculated using Eqs.~(\ref{eq:si:optError}) and (\ref{eq:si:BogCorr}). When the input optical field is in the vacuum state, the result is
\begin{equation}\label{eq:si:optErrorVar}
\langle\abs{\hat{\varepsilon}}^2\rangle=\zeta_1^2\int_{0}^{T}\left[B_1^2(t)+B_2^2(t)\right]dt+\frac{1-\zeta_1^2}{2}\int_{0}^{T}[B_1(t)-B_2(t)]^2dt.
\end{equation}
To make the minimization of this expression more transparent, it can be transformed using the identities (see Sec.~\ref{sec:si:identities} for details)
\begin{align}\label{eq:si:idents-SerTel}
&\int_{0}^{T}B_1^2(t)dt=\frac{1}{4\zeta_1^2}\int_{0}^{T}\tilde{f}^2(t)dt + \frac{M_1^2}{2\zeta_1},
&\int_{0}^{T}B_2^2(t)dt=\frac{1}{4\zeta_1^2}\int_{0}^{T}\tilde{f}^2(t)dt + \frac{M_2^2}{2\zeta_1},
\end{align}
into
\begin{equation}\label{eq:si:optErrorVar2}
\langle\abs{\hat{\varepsilon}}^2\rangle=\frac{1}{2}\int_{0}^{T}\tilde{f}^2(t)dt+\zeta_1 M^2+\frac{1-\zeta_1^2}{2}\int_{0}^{T}[B_1(t)-B_2(t)]^2dt.
\end{equation}
From this expression it is already clear that if $\zeta_1>0$, the teleportation error cannot be arbitrarily small; this fact agrees well with the intuition that a good teleportation protocol requires the interaction between Alice (the optical field) and Bob (oscillator 1) to be predominantly of the two-mode-squeezing rather than beam-splitter type.

It can be shown that if the only decoherence channel for the localized oscillators is the traveling field eventually measured by our detectors, as assumed throughout this work, making both measurement rates time-dependent is actually never required. This is because we can always consider the dynamics in the interaction time frame $\tau_j$ of one of the oscillators (in which its rate appears constant); we therefore fix $\Gamma_1$ (and hence the optical damping $\gamma_1$) to be time-independent (in the interval $[0,T]$) without loss of generality.

To find the minimum teleportation error, the expression~(\ref{eq:si:optErrorVar2}) subject to the constraints~(\ref{eq:si:M-constraint_seq}) can be variationally optimized. Instead of approaching it with the regular procedure, we here present a simpler argument consisting of two steps. 
First, we minimize only the first term $(1/2)\int_0^T\tilde{f}^2(t)dt$ under the constraint $M_1=1$ using variational calculus. This is an easily solvable problem, because this term does not depend on the measurement rates $\Gamma_j$, and the result is
\begin{equation}\label{eq:si:telepftilde}
\tilde{f}_{2\to 1}(t)=\frac{\zeta_1\sqrt{\Gamma_1}}{\sinh(\gamma_1 T/2)}e^{-\gamma_1 t/2}.
\end{equation}
Second, we observe that if for $\tilde{f}(t)=\tilde{f}_{2\to 1}(t)$ there exists a $\Gamma_2(t)$ such that the third term in \eqref{eq:si:optErrorVar2} reaches the smallest value allowed mathematically, namely zero (recall that $B_j(t)\in\mathbb{R}$ and $\zeta_j\in [-1,1]$), while $M_2=1$ is fulfilled, then this must be the desired minimum. Indeed this turns out to be the case, as we will now demonstrate by finding the time-dependent measurement rate $\Gamma_2(t)$ such that for all $t\in [0,T]$
\begin{equation}\label{eq:si:B-equal_seqTel}
B_1(t)=B_2(t) \Leftrightarrow \sqrt{\Gamma_1}M_1(t)=\sqrt{\Gamma_2(t)}M_2(t),
\end{equation}
where the equivalence follows from Eqs.~(\ref{eq:si:BSerTel}) and the present assumption of $\Gamma_1(t)=\Gamma_1$. 
Taking the time derivative of \eqref{eq:si:B-equal_seqTel} and using Eqs.~(\ref{eq:si:M-t_seq}) results in the following non-linear differential equation for $\Gamma_2(t)/\Gamma_1$
\begin{equation}\label{eq:si:Gamma-diffEq-SerTel}
  \frac{1}{\gamma_1}\frac{d}{dt}\left(\frac{\Gamma_2(t)}{\Gamma_1}\right) = \frac{\Gamma_2(t)}{\Gamma_1}\left[1 + \frac{\Gamma_2(t)}{\Gamma_1}\right]\coth(\gamma_1t/2).
\end{equation}
The particular solution to this differential equation that satisfies \eqref{eq:si:B-equal_seqTel} is 
\begin{equation}\label{eq:si:OptGamma2SerTelep}
\frac{\Gamma_2(t)}{\Gamma_1}=\frac{\sinh^2(\gamma_1 t/2)}{\sinh^2(\gamma_1 T/2)-\sinh^2(\gamma_1 t/2)}.
\end{equation}
That this is indeed the desired particular solution can be verified, e.g., by confirming that \eqref{eq:si:B-equal_seqTel} is satisfied at the endpoint $t=T$ (seeing as $\Gamma_2(t)\xrightarrow{t\rightarrow T}\infty$ and $M_2(T)=0$, l'H\^{o}pital's rule must be invoked).
Finally, we note that Eqs.~(\ref{eq:si:B-equal_seqTel}) and (\ref{eq:si:idents-SerTel}) imply $M^2_2=M^2_1$ and, in view of $M_j\geq 0$ as follows from Eqs.~(\ref{eq:si:M-t_seq}), we have $M_2=1$; thus the combination of $\tilde{f}(t)$ and $\Gamma_2(t)$ we have arrived at provides the global minimum of the error variance~(\ref{eq:si:optErrorVar2}) subject to the constraints~(\ref{eq:si:M-constraint_seq}). The corresponding minimum error variance is given by
\begin{equation}
\langle\abs{\hat{\varepsilon}}^2\rangle_{2\to 1}=\frac{\zeta_1}{1-e^{-\gamma_1 T}},
\end{equation}
which, remarkably, is independent of $\zeta_2$ that characterizes the interaction type of Charlie's oscillator.
Using this variance, we define the state-transfer fidelity as
\begin{equation}
    \mathcal{F}\equiv 1/(1+\langle\abs{\hat{\varepsilon}}^2\rangle),
\end{equation}
which coincides with the state-overlap fidelity for an arbitrary coherent input state~\cite{hammerer_quantum_2010}.

As was mentioned before, these results are easily generalized to the case when the input optical field is in the thermal state with photon occupancy $n_\t{in}$. In this case, the error given by \eqref{eq:si:optErrorVar2} has to be multiplied by the constant $\left(2n_\t{in}+1\right)$, which leaves the minimization procedure outlined above completely unchanged. The found optimum filter $\tilde{f}(t)$ and the time dependence of the drive $\Gamma_2(t)$ are valid for any $n_\t{in}$.
The minimum error is given by $\langle\abs{\hat{\varepsilon}}^2\rangle_{2\to 1}=\left(2n_\t{in}+1\right)\zeta_1/(1-e^{-\gamma_1 T})$.

As a concluding remark, the optimum filter $\tilde{f}(t)$ is applied to the renormalized measurement record $\tilde{m}(t)$, which may not be convenient from a practical perspective. By inverting the renormalization that led to $\tilde{f}(t)$, we can obtain the filtering function $f(t)$ that can be applied to the raw measurement record $m(t)$ to the same effect, $\int_{0}^{T}\tilde{f}(t)\,\tilde{m}(t)dt=\int_{0}^{T}f(t)\,m(t)dt$. The inversion is done via
\begin{equation}
f(t)=\tilde{f}(t)+(\zeta_1+\zeta_2)\sqrt{\Gamma_2(t)}M_2(t),
\end{equation}
and using the expression from \eqref{eq:si:telepftilde} along with \eqref{eq:si:B-equal_seqTel}, we find
\begin{equation}
f_{2\to 1}(t)=\frac{\sqrt{\Gamma_1}}{2\sinh(\gamma_1 T/2)}\left( (\zeta_1+\zeta_2)e^{\gamma_1 t/2}+(\zeta_1-\zeta_2)e^{-\gamma_1 t/2}\right).
\end{equation}

\subsection{The sequential case: conditional direct transfer}\label{sec:si:SeqDir}

Reversing the direction of the state transfer in the sequential case from $2\to 1$ to $1\to 2$ requires only minor modifications to the steps taken to solve the $2\to 1$ teleportation problem; the crucial difference is that now the relevant pretrodiction observable is $\hat{b}_{I,1}(0)+\hat{b}_{I,2}(T)$. The measurement record is again renormalized according to 
\eqref{eq:si:measRecRenorm}, but this time with
 $\phi_m(t)$ given by
\begin{equation}
\phi_m(t)=-(\zeta_1+\zeta_2)e^{-\tilde{\tau}_2(t)/2}
\int_{t}^{T}e^{\tilde{\tau}_2(t')/2}\sqrt{\Gamma_2(t')}\,m(t')\,dt'.
\end{equation}
The measurement record is processed by a filter function $\tilde{f}(t)$ to extract the pretrodiction measurement result,
\begin{equation}\label{eq:si:mathcalM-SerDir}
\mathcal{M}=\int_{0}^{T}\tilde{f}(t)\tilde{m}(t)dt=
\hat{\varepsilon} - M_1 \hat{b}_{I,1}(0) - M_2 \hat{b}_{I,2}(T).
\end{equation}
Again, $\hat{\varepsilon}$ is the optical error operator. The partial (time-dependent) transfer coefficients for conditional direct transfer are given by [cf.\ Eqs.~(\ref{eq:si:M-t_seq})]
\begin{subequations}\label{eq:si:M-t_seqDir}
\begin{align}
M_1(t)&=\int_{t}^{T}\tilde{f}(t')\sqrt{\Gamma_1(t')}\,e^{[\tau_1(t)-\tau_1(t')]/2}dt',\\
M_2(t)&=\int_{0}^{t}\tilde{f}(t')\sqrt{\Gamma_2(t')}\,e^{[\tilde{\tau}_2(t)-\tilde{\tau}_2(t')]/2}dt',
\end{align}
\end{subequations}
in terms of which the final transfer coefficients $M_1$ and $M_2$ are now given by [cf.\ Eqs.~(\ref{eq:si:M-final-SerTel})]
\begin{align}
&M_1\equiv M_1(0),&M_2\equiv M_2(T).
\end{align}
As previously, the error operator $\hat{\varepsilon}$ is expressed in the form~(\ref{eq:si:optError}),
but this time with new temporal noise modes $B_j(t)$,
\begin{align}\label{eq:si:BSerDir}
&B_1(t)=\frac{\tilde{f}(t)}{2\zeta_1}-\sqrt{\Gamma_1(t)}M_1(t),&
&
B_2(t)=\frac{\tilde{f}(t)}{2\zeta_1}-\sqrt{\Gamma_2(t)}M_2(t).&
\end{align}
The expression for the single-quadrature variance of the error in \eqref{eq:si:optErrorVar} remains valid, now in terms of the new $B_j(t)$ [Eqs.~(\ref{eq:si:BSerDir})], however the identities that transform the variance of the error to a simplified form analogous to \eqref{eq:si:optErrorVar2} are different (see Sec.~\ref{sec:si:identities} for details),
\begin{align}\label{eq:si:idents-SerDir}
&\int_{0}^{T}B_1^2(t)dt=\frac{1}{4\zeta_1^2}\int_{0}^{T}\tilde{f}^2(t)dt - \frac{M_1^2}{2\zeta_1},
&\int_{0}^{T}B_2^2(t)dt=\frac{1}{4\zeta_1^2}\int_{0}^{T}\tilde{f}^2(t)dt - \frac{M_2^2}{2\zeta_1}.
\end{align}
The simplified form itself is different from \eqref{eq:si:optErrorVar2} only by the sign of $\zeta_1$,
\begin{equation}
\langle\abs{\hat{\varepsilon}}^2\rangle=\frac{1}{2}\int_{0}^{T}\tilde{f}^2(t)dt-\zeta_1 M^2+\frac{1-\zeta_1^2}{2}\int_{0}^{T}[B_1(t)-B_2(t)]^2 dt.
\end{equation}
Hence, the optimization procedure presented in Sec.~\ref{sec:si:SeqTel} carries over to the present case. 

The optimum filter for the direct transfer that can be applied to the renormalized measurement record is
\begin{equation}
\tilde{f}_{1\to 2}(t)=\frac{\zeta_1\sqrt{\Gamma_1}}{\sinh(\gamma_1 T/2)}e^{\gamma_1 (T-t)/2},
\end{equation}
and the filter for the bare measurement record is found using
\begin{equation}
f(t)=\tilde{f}(t)-(\zeta_1+\zeta_2)\sqrt{\Gamma_2(t)}M_2(t),
\end{equation}
to be
\begin{equation}
f_{1\to 2}(t)=\frac{\sqrt{\Gamma_1}}{2\sinh(\gamma_1 T/2)}\left((\zeta_1+\zeta_2)e^{\gamma_1 (t-T)/2}+(\zeta_1-\zeta_2)e^{-\gamma_1 (t-T)/2}\right).
\end{equation}
The optimum time-dependent $\Gamma_2(t)$, which is found from the condition $B_1(t)=B_2(t)$, obeys the differential equation
\begin{equation}
  \frac{1}{\gamma_1}\frac{d}{dt}\left(\frac{\Gamma_2(t)}{\Gamma_1}\right) = \frac{\Gamma_2(t)}{\Gamma_1}\left[1 + \frac{\Gamma_2(t)}{\Gamma_1}\right]\coth(\gamma_1[t-T]/2).
\end{equation}
and is given by
\begin{equation}\label{eq:si:OptGamma2SerDirTrans}
\frac{\Gamma_2(t)}{\Gamma_1}=\frac{\sinh^2(\gamma_1 [T-t]/2)}{\sinh^2(\gamma_1 T/2)-\sinh^2(\gamma_1 [T-t]/2)}.
\end{equation}

\subsection{The parallel case: teleportation}\label{sec:si:parallel}

In the parallel configuration, two oscillators interact with two independent input optical fields, described by their Bogoliubov operators $\hat{u}^{\zeta_1}_{\t{in},1}$ and $\hat{u}^{\zeta_2}_{\t{in},2}$. The state transfer problem is symmetric with respect to the direction, $2\to 1$ or $1\to 2$, and the state transfer is accomplished via teleportation in both cases. We choose the direction $2\to 1$, and restrict ourselves to the case when $\zeta_2=-\zeta_1$. It can be shown that perfect teleportation is impossible in the parallel configuration if $\zeta_2\neq-\zeta_1$. The equations of motion in the frame rotating with the frequency $\Omega$ are
\begin{align}
\frac{d}{dt}\hat{b}_{I,1}(t)&=-\frac{\gamma_1(t)}{2}\hat{b}_{I,1}(t)-i\sqrt{\Gamma_1(t)}\,\hat{u}^{\zeta_1}_{\t{in},1}(t),\\
\frac{d}{dt}\hat{b}_{I,2}(t)&=-\frac{\gamma_2(t)}{2}\hat{b}_{I,2}(t)-i\sqrt{\Gamma_2(t)}\,\hat{u}^{\zeta_2}_{\t{in},2}(t),
\end{align}
After having interacted with the oscillators, the two fields are combined on a 50:50 beamsplitter and homodyne detected with two independent local oscillators. One homodyne is set to detect the amplitude quadrature, and the other to detect the phase quadrature of the field. The two photocurrents are demodulated at the oscillator frequency $\Omega$ to produce two complex measurement records, $m_1(t)$ and $m_2(t)$, which are then combined into the more convenient variables $m_{-}$ and $m_{+}$ given by
\begin{align}
m_{-}(t)=\frac{1}{\sqrt{2}}\left(m_1(t)-\frac{i}{\zeta_1}m_2(t)\right)&=\frac{1}{2\zeta_1}\left(-i\hat{u}^{\zeta_1}_{\t{in},1}(t)+i\hat{u}^{-\zeta_1}_{\t{in},2}(t)\right)-\sqrt{\Gamma_1(t)}\hat{b}_{I,1}(t)-\sqrt{\Gamma_2(t)}\hat{b}_{I,2}(t),\\
m_{+}(t)=\frac{1}{\sqrt{2}}\left(m_1(t)+\frac{i}{\zeta_1}m_2(t)\right)&=\frac{1}{2\zeta_1}\left(-i\hat{u}^{\zeta_1}_{\t{in},2}(t)+i\hat{u}^{-\zeta_1}_{\t{in},1}(t)\right).
\end{align}
The measurement records are integrated over the interval $[0,T]$, the first one weighted using the filter function $f(t)$ and the second one with $g(t)$, and added together to provide the pretrodiction measurement result $\mathcal{M}$,
\begin{equation}\label{eq:si:mathcalM-par}
\mathcal{M}=\int_{0}^{T}f(t)m_{-}(t)dt+\int_{0}^{T}g(t)m_{+}(t)dt = \hat{\varepsilon} - M_1 \hat{b}_{I,1}(T) - M_2 \hat{b}_{I,2}(0),
\end{equation}
where the final transfer coefficients $M_j$ and the functions $M_j(t)$ they derive from are given by Eqs.~(\ref{eq:si:M-final-SerTel}) and (\ref{eq:si:M-t_seq}), as in the sequential teleportation case [note that the present assumption $\zeta_2=-\zeta_1$ implies $\tilde{\gamma}_2(t)=\gamma_2(t)$ and $\tilde{\tau}_2(t)=\tau_2(t)$, see Eqs.~(\ref{eq:si:tilde2-defs})]. 
The measurement error is given by
\begin{equation}
\hat{\varepsilon}=-i\int_{0}^{T}B_1(t)\hat{u}^{\zeta_1}_{\t{in},1}(t)dt
+i\int_{0}^{T}B_2(t)\hat{u}^{-\zeta_1}_{\t{in},2}(t)dt
+\int_{0}^{T}\frac{g(t)}{2\zeta_1}\left(-i\hat{u}^{\zeta_1}_{\t{in},2}(t)+i\hat{u}^{-\zeta_1}_{\t{in},1}(t)\right)dt,
\end{equation}
where the $B_j(t)$ are defined as in Eqs.~(\ref{eq:si:BSerTel}). The variance of the error is found to be
\begin{equation}\label{eq:si:optError-Par1}
\langle\abs{\hat{\varepsilon}}^2\rangle=\frac{2\zeta_1^2}{1+\zeta_1^2}\int_{0}^{T}\left[B_1^2(t)+B_2^2(t)\right]dt+\frac{1+\zeta_1^2}{2}
\int_{0}^{T}\left[\left(\frac{g(t)}{2\zeta_1}-\frac{1-\zeta_1^2}{1+\zeta_1^2}B_1(t)\right)^2+\left(\frac{g(t)}{2\zeta_1}-\frac{1-\zeta_1^2}{1+\zeta_1^2}B_2(t)\right)^2\right] dt.
\end{equation}
This variance is to be minimized over $f(t)$, $g(t)$, and $\Gamma_2(t)$ under the constraints $M_1=M_2=1$. For given $f(t)$ and $\Gamma_2(t)$, the minimum over $g(t)$, which is unconstrained, is straightforwardly found to be achieved when 
$\frac{g(t)}{2\zeta_1}=\frac{1-\zeta_1^2}{1+\zeta_1^2}\frac{B_1(t)+B_2(t)}{2}$. 
Inserting this in \eqref{eq:si:optError-Par1} and using that the identities~(\ref{eq:si:idents-SerTel}) also hold in the parallel case (see Sec.~\ref{sec:si:identities} for details), we find
\begin{equation}\label{eq:si:optError-Par2}
\langle\abs{\hat{\varepsilon}}^2\rangle= \frac{1}{1+\zeta_1^2}\int_0^T f^2(t) dt + \frac{\zeta_1}{1+\zeta_1^2}(M_1^2 + M_2^2) + \frac{1}{4}\frac{(1-\zeta_1^2)^2}{1+\zeta_1^2}\int_0^T [B_1(t) - B_2(t)]^2 dt.
\end{equation}
In view of the similarity of \eqref{eq:si:optError-Par2} to the analogous \eqref{eq:si:optErrorVar2} for the sequential teleportation case, including the definitions of $M_j$ and $B_j(t)$, we can employ the same optimization strategy as in Sec.~\ref{sec:si:SeqTel}. 
It follows that the optimal filter $f(t)$ equals that of the sequential teleportation case, $f(t) = \tilde{f}_{2\to 1}(t)$ [\eqref{eq:si:telepftilde}], and that the optimal time-dependent measurement rate of the second oscillator $\Gamma_2(t)$ is given by \eqref{eq:si:OptGamma2SerTelep}. 
Invoking again $B_1(t)=B_2(t)$ along with the constraint $M_1=M_2=M$, it is clear from comparison of Eqs.~(\ref{eq:si:optError-Par2}) and (\ref{eq:si:optErrorVar2}) that the minimized noise variance in the parallel case equals that of the sequential case times a factor of $2/(1+\zeta_1^2)$,
\begin{equation}
\langle\abs{\hat{\varepsilon}}^2\rangle_\t{par}=\frac{2\zeta_1/(1+\zeta_1^2)}{1-e^{-\gamma_1 T}}.
\end{equation}

\subsection{The unconditional direct state transfer}\label{sec:si:uncond}

To find the error of the unconditional state transfer from oscillator 1 to oscillator 2, we return to Eqs.~(\ref{eq:si:b1}) and (\ref{eq:si:b2}) describing the evolution of the two oscillators sequentially coupled to a field, and solve their unconditional evolution. 
The coupling term that enables unconditional transfer is the 2nd term on the right-hand side of \eqref{eq:si:b2} --- exactly the term that we found it convenient to eliminate by renormalization in the optimization of the conditional sequential schemes. 
The rotating-frame annihilation operator of oscillator 2 is expressed at the final time as the function of the input light field and the initial annihilation operators of both oscillators. The unconditional state transfer error is defined by
\begin{equation}\label{eq:si:errorOpUncond1}
\hat{\varepsilon}_\t{uc}=\hat{b}_{I,2}(T)+\hat{b}_{I,1}(0);
\end{equation}
note that, as a matter of convention, we are here considering the transfer of $-\hat{b}_{I,1}(0)$ into $\hat{b}_{I,2}(T)$. 
Without using the measurement record, there is no way to completely eliminate the information about the initial states of the two oscillators from $\hat{\varepsilon}_\t{uc}$. To circumvent this complication, we assume that oscillators 2 and 1 both are in the vacuum state in the beginning. 
This is a meaningful way of lower-bounding the variance of $\hat{\varepsilon}$ compared to any bigger family of initial states (including the infinite families assumed in optimizing the conditional protocols). The contributions of the initial states decay as the interaction time increases, so our assumptions are inconsequential for the limits of the error variance as $T\to\infty$.

The rotating-frame annihilation operator of the first oscillator is found by solving \eqref{eq:si:b1} forward in time,
\begin{equation}\label{eq:si:b1-sol-uncond}
\hat{b}_{I,1}(t)=e^{-\tau_1(t)/2}\hat{b}_{I,1}(0)-
ie^{-\tau_1(t)/2}\int_{0}^{t}e^{\tau_1(t')/2}\sqrt{\Gamma_1(t')}\,\hat{u}^{\zeta_1}_\t{in}(t')dt'.
\end{equation}
Plugging this into \eqref{eq:si:b2}
for the annihilation operator of the second oscillator and solving for the final time, we find
\begin{equation}\label{eq:si:b2-sol-uncond}
\hat{b}_{I,2}(T)=K_2\hat{b}_{I,2}(0)-K_1\hat{b}_{I,1}(0)-i\int_{0}^{T}C_2(t)\,\hat{u}^{\zeta_2}_\t{in}(t)dt+i\int_{0}^{T}C_1(t)\,\hat{u}^{\zeta_1}_\t{in}(t)dt,
\end{equation}
where, in terms of $\tau_j(t)$ [\eqref{eq:si:tau-def}], we have introduced 
\begin{align}\label{eq:si:b2-sol-Uncond-defs}
K_2&=e^{-\tau_2(T)/2},\\
K_1&=(\zeta_1+\zeta_2)\int_{0}^{T}e^{[\tau_2(t')-\tau_2(T)]/2}\sqrt{\Gamma_1(t')\Gamma_2(t')}\,e^{-\tau_1(t')/2}dt',\\
C_2(t)&=e^{[\tau_2(t)-\tau_2(T)]/2}\sqrt{\Gamma_2(t)},\\
C_1(t)&=(\zeta_1+\zeta_2)e^{\tau_1(t)/2}\sqrt{\Gamma_1(t)}\int_{t}^{T}e^{[\tau_2(t')-\tau_2(T)]/2}\sqrt{\Gamma_1(t')\Gamma_2(t')}\,e^{-\tau_1(t')/2}dt'.
\end{align}
The total error~(\ref{eq:si:errorOpUncond1}) can now be expressed from \eqref{eq:si:b2-sol-uncond} as a sum of contributions from the optical field, the initial state of oscillator 1, and the initial state of oscillator 2,
\begin{equation}
\hat{\varepsilon}_\t{uc}=\hat{\varepsilon}_\t{o}+\hat{\varepsilon}_{b1}+\hat{\varepsilon}_{b2},
\end{equation}
where the optical contribution is given by
\begin{equation}
\hat{\varepsilon}_\t{o}=-i\int_{0}^{T}C_2(t)\,\hat{u}^{\zeta_2}_\t{in}(t)dt+i\int_{0}^{T}C_1(t)\,\hat{u}^{\zeta_1}_\t{in}(t)dt,
\end{equation}
and the other two by
\begin{align}
&\hat{\varepsilon}_{b1}=(1-K_1)\hat{b}_{I,1}(0),
&\hat{\varepsilon}_{b2}=K_2\hat{b}_{I,2}(0).
\end{align}
The variance of the optical error is
\begin{equation}
\langle|\hat{\varepsilon}_\t{o}|^2\rangle=\frac{1+\zeta_2^2}{2}\int_{0}^{T}C_2^2(t)dt+\frac{1+\zeta_1^2}{2}\int_{0}^{T}C_1^2(t)dt-(1+\zeta_1\zeta_2)\int_{0}^{T}C_1(t)C_2(t)dt.
\end{equation}
Using the identities (see Sec.~\ref{sec:si:identities} for details)
\begin{align}\label{eq:si:idents-uncond}
&\int_{0}^{T}C_2^2(t)dt=\frac{1}{2\zeta_2}\left(1-K_2^2\right),
&\int_{0}^{T}C_1^2(t)dt=\frac{\zeta_1+\zeta_2}{\zeta_1}\int_{0}^{T}C_1(t)C_2(t)dt-\frac{K_1^2}{2\zeta_1},
\end{align}
the variance of the optical error is re-expressed as
\begin{equation}
\langle|\hat{\varepsilon}_\t{o}|^2\rangle=\frac{1+\zeta_2^2}{4\zeta_2}\left(1-K_2^2\right)-\frac{1+\zeta_1^2}{4\zeta_1}K_1^2+\frac{(1-\zeta_1^2)(\zeta_2-\zeta_1)}{2\zeta_1}\int_{0}^{T}C_1(t)C_2(t)dt.
\end{equation}

In the main text, the error of the unconditional state transfer is evaluated for the time-dependent readout rate realizing the \emph{conditional} optimum (hence, we do not attempt an optimization of the unconditional scheme). As a matter of convenience, we chose to fix $\Gamma_2(t)=\Gamma_2$ and vary $\Gamma_1(t)$, which is then given by
\begin{equation}\label{eq:si:Gamma1VarUncond}
\Gamma_1(t)=\Gamma_2\frac{\sinh^2(\zeta_1\Gamma_2 t)}{\sinh^2(\zeta_1\Gamma_2 T)-\sinh^2(\zeta_1\Gamma_2 t)}.
\end{equation}
In the limit $T\to\infty$ for $\Gamma_1(t)$ given by \eqref{eq:si:Gamma1VarUncond} and for $\zeta_{1,2}>0$, we have $K_2\to 0$ and $K_1\to 1$. The variance of the full error of the unconditional state transfer in this case is
\begin{equation}
\langle|\hat{\varepsilon}_\t{uc}|^2\rangle_{T\to\infty}=\frac{\zeta_1-\zeta_2}{2}\left(\frac{1-\zeta_1\zeta_2}{2\zeta_1\zeta_2}-\frac{1-\zeta_1^2}{\zeta_1}\int_{0}^{T}C_1(t)C_2(t)dt\right).
\end{equation}
When $\zeta_1=\zeta_2>0$, the unconditional error is zero at $T\to\infty$. To show that $K_1\to 1$ for $T\to\infty$, we calculate the interaction time of the first oscillator,
\begin{equation}
\tau_1(t)=\int_{0}^{t}2\zeta_1\Gamma_1(t')dt'=-2\zeta_1\Gamma_2 t+\tanh\left(\zeta_1\Gamma_2T\right)\ln\left(\frac{\sinh[\zeta_1\Gamma_2(T+t)]}{\sinh[\zeta_1\Gamma_2(T-t)]}\right).
\end{equation}
In the limit $T\to\infty$, we find that $\sqrt{\Gamma_1(t)\,\Gamma_2}e^{-\tau_1(t)/2}\approx 2\Gamma_2e^{-\zeta_1\Gamma_2T}\sinh(\zeta_1 \Gamma_2 t)$, and therefore
\begin{equation}
K_1\approx 2(\zeta_1+\zeta_2)e^{-(\zeta_1+\zeta_2)\Gamma_2T}\int_{0}^{T}e^{\zeta_2\Gamma_2 t}\sinh(\zeta_1 \Gamma_2 t)\Gamma_2dt\to 1.
\end{equation}

\subsection{Comparison of direct state transfer analysis with Ref.~\cite{jahne_high-fidelity_2007}}\label{sec:si:comp-direct}

In Ref.~\cite{jahne_high-fidelity_2007}, the authors solved the problem of perfectly catching a wavepacket that leaks out of one optical cavity by another in a situation when the reflectivity of the transmitting mirror of the first cavity can be time-dependent. The optimum time-dependence of the coupling rate of the first cavity was found to be
\begin{equation}\label{eq:si:JahneOptimumGamma}
\Gamma_1(t)=\Gamma_2\left/\left(e^{2\Gamma_2(T-t)}-1\right)\right. .
\end{equation}
The situation is also covered by our analysis and corresponds to the case when $\zeta_1=\zeta_2=1$. The optimum time dependence of $\Gamma_1$ given by \eqref{eq:si:JahneOptimumGamma} agrees with our result in \eqref{eq:si:Gamma1VarUncond} in the limit of large interaction strength, when $t\to T$ and $\Gamma_2 T\gg 1$, in which case both of them can be approximated as $\Gamma_1(t)\approx 1/\big(2(T-t)\big)$.
At finite interaction strengths, the two expressions are not exactly the same because of the different definitions of the state transfer fidelity that were used to obtain them.

\subsection{Identities linking mode function integrals and transfer coefficients based on unitarity}\label{sec:si:identities}

In analyzing the various state transfer schemes, the calculations were simplified by using certain identities linking integrals over temporal mode functions with the final transfer coefficients. In the conditional schemes, the identities~(\ref{eq:si:idents-SerTel}) and (\ref{eq:si:idents-SerDir}) linked $\int_0^T B_j^2(t) dt$ and $\int_0^T \tilde{f}^2(t) dt$ to $M_j^2$, whereas in the unconditional scheme, the identities~(\ref{eq:si:idents-uncond}) linked $\int_0^T C_i(t)C_j(t) dt$ to $K_j^2$.
In the present section, we indicate how these can be derived from unitarity.

To derive the identities for the conditional schemes, two key observations are that the quadratures of the measurement operator $\mathcal{M}$ commute $[\mathcal{M}+\mathcal{M}^\dagger,\mathcal{M}-\mathcal{M}^\dagger]=0\Leftrightarrow [\mathcal{M},\mathcal{M}^\dagger]=0$, as follows from \eqref{eq:si:m-general}, and that initial-time annihilation operators $\hat{b}_{I,j}(0)$ for the localized oscillators commute with all input field operators $\hat{u}^\zeta_\t{in}(t)$ whereas final-time operators $\hat{b}_{I,j}(T)$ commute with all output operators $\hat{u}^\zeta_\t{out}(t)$. 

Considering here the sequential teleportation case as an example, 
we rewrite the measurement operator~(\ref{eq:si:mathcalM-seq}) using \eqref{eq:si:optError} so as to group the initial- and final-time annihilation operators with the field operators with which they commute,
\begin{equation}\label{eq:si:mathcalM-ident}
\mathcal{M} + M_1 \hat{b}_{I,1}(T) = - M_2 \hat{b}_{I,2}(0)  -i\int_{0}^{T}B_1(t)\hat{u}^{\zeta_1}_\t{in}(t)dt
+i\int_{0}^{T}B_2(t)\hat{u}^{-\zeta_1}_\t{in}(t)dt.
\end{equation}
We now calculate the commutator $[\hat{O},\, \hat{O}^\dagger]$ twice where $\hat{O}$ is either the left- or right-hand side of \eqref{eq:si:mathcalM-ident}; using the observations above along with \eqref{eq:si:BogComm} and the bosonic same-time commutation relation $[\hat{b}_{I,j}(t),\, \hat{b}_{I,j}^\dagger(t)]=1$, this results in the identity
\begin{align}\label{eq:si:mathcalMcomm1}
    M_1^2 = M_2^2 + 2\zeta_1 \int_0^T [B_1^2(t) - B_2^2(t)]dt.
\end{align} 
The desired identities~(\ref{eq:si:idents-SerTel}) for the individual oscillators follow from \eqref{eq:si:mathcalMcomm1} by evaluating it for either $\Gamma_2=0$ or $\Gamma_1=0$ while recalling the definitions~(\ref{eq:si:M-t_seq}) and (\ref{eq:si:BSerTel}). Applying the same procedure to the conditional direct transfer results in \eqref{eq:si:mathcalMcomm1} with the $M_{1,2}^2$ terms swapped, seeing as the role of initial- and final-time states for oscillators 1 and 2 is now inverted; accordingly, this entails the identities~(\ref{eq:si:idents-SerDir}) for the individual oscillators in which the signs of $M_{1,2}^2$ are flipped relative to the analogous teleportation expressions~(\ref{eq:si:idents-SerTel}).

Turning now to the unconditional calculation, the key observation is the conservation of the commutation relation $[\hat{b}_{I,2}(T),\hat{b}_{I,2}^\dagger(T)]=1$. Evaluating $[\hat{b}_{I,2}(T),\hat{b}_{I,2}^\dagger(T)]_{\Gamma_1=0}=1$ using \eqref{eq:si:b2-sol-uncond} yields the first of Eqs.~(\ref{eq:si:idents-uncond}), while the evaluation of $[\hat{b}_{I,2}(T),\hat{b}_{I,2}^\dagger(T)]-[\hat{b}_{I,2}(T),\hat{b}_{I,2}^\dagger(T)]_{\Gamma_1=0}=0$ yields the second.

\subsection{The effect of finite measurement rate}\label{sec:si:finiteMeasRate}
The expressions for the optimum variations of the measurement rates given by Eqs.~(\ref{eq:si:OptGamma2SerTelep}) and (\ref{eq:si:OptGamma2SerDirTrans}) diverge as $t\to T$ and $t\to 0$, respectively. In a practical implementation of the state transfer protocols, the measurement rates would need to stay below a maximum value, dictated by the experimental limitations, and the requirement to stay within the validity region of the rotating wave approximation. To show the effect that a limitation on the measurement rate would have, in \figref{fig:bounds}b of the main manuscript we evaluated the added noise in the sequential teleportation protocol for a non-diverging drive of the second oscillator. We choose the expression for the drive to be a truncated version of the optimum expression in Eqs.~(\ref{eq:si:OptGamma2SerTelep}), given by   
\begin{equation}\label{eq:si:OptGamma2SerTelepTrunc}
\frac{\Gamma_2(t)}{\Gamma_1}=\min\left[\frac{\sinh^2(\alpha \Gamma_1 t)}{\sinh^2(\alpha \Gamma_1 T/2)-\sinh^2(\alpha \Gamma_1 t/2)},\; \Gamma_{2,\t{max}}\right],
\end{equation}
where $\zeta_1$ additionally was replaced with a free parameter $\alpha$, which is adjusted to satisfy the requirement $M_1=M_2$. The value of $\alpha$ is close to $\zeta_1$. The truncation point is characterized by the ratio $r_\t{max}=\Gamma_{2,\t{max}}/\Gamma_1$ in \figref{fig:bounds}b. The expression in \eqref{eq:si:OptGamma2SerTelepTrunc}, in general, is not optimal, which means that further optimization of the state transfer protocols is possible provided the experimental constraints (see \cite{jahne_high-fidelity_2007,korotkov_flying_2011} for the discussion of a similar issue with direct unconditional transfer).

\end{document}